\documentclass[12pt]{iopart}

\usepackage{graphicx}
\def\esy{E_{sym}}

\begin{document}

\title[Probing the symmetry energy at high baryon density with heavy ion collisions]{PROBING THE SYMMETRY ENERGY
\\AT HIGH BARYON DENSITY WITH HEAVY ION COLLISIONS}

\author{V. Greco, M. Colonna, M. Di Toro}

\address{Department of Physics and Astronomy, University of Catania, and INFN-LNS, Via. S. Sofia 62\\
Catania, I-95125,
Italy\\
greco@lns.infn.it}
\ead{greco@lns.infn.it}

\author{H.H. Wolter}

\address{Fakultat fur Physik, Ludwig-Maximilians- Universitat Munchen\\
D-85748 Garching, Germany\\
}

\begin{abstract}
The nuclear symmetry energy at densities above saturation density ($\rho_0\sim 0.16 fm^{-3}$) is poorly
constrained theoretically and very few relevant experimental data exist. Its study is possible through Heavy Ion Collisions (HIC) at energies $E/A> 200$ MeV, particularly with beams of neutron-rich radioactive nuclei.
The energy range implies that the momentum dependence
of the isospin fields, i.e. the difference of the effective masses on protons and neutrons, also has  to be investigated before a safe constraint on $\esy(\rho)$ is possible. We discuss the several observables which have been suggested, like $n/p$ emission and their collective flows and the ratio of meson yields with different isospin projection, $\pi^-/\pi^+$ and $K^0/K^+$. We point out several physical mechanisms that should
be included in the theoretical models to allow a direct comparison to the more precise experiments
 which will be able to distinguish the isospin projection of the detected particles:
CSR/Lanzhou, FAIR/GSI, RIBF/RIKEN, FRIB/MSU.

\end{abstract}

\maketitle

\section{Introduction}
The study of the equation of state (EOS) of nuclear matter has been one of the main challenges of
nuclear physics in the last 30 years. The main strategy for its determination has been the investigation
of the pletheora of data from heavy ion experiments in a wide range of beam energies $E/A\simeq 0.02-2$ GeV.
In particular, through the analyisis of collective flows of nucleons and of strange meson
production it has been possible to constrain the EOS in a wide range of densities up to $\rho \sim 2 \rho_0$ \cite{Danielewicz:2002pu,Fuchs:2005yn}. This endeavor has been performed mainly with respect to the properties of symmetric nuclear
matter with an equal number of protons and neutrons. However, it is known already from the Bethe-Weizs\"acker mass
formula that the difference in the neutron-proton content gives a contribution to the energy
called the symmetry energy, i.e. the energy arising from the breaking of the neutron-proton symmetry (apart form the Coulomb energy). The symmetry energy $E_{sym}$ can be defined from the following expansion of the energy density $\epsilon(\rho,\rho_3) \equiv \epsilon(\rho)+\rho E_{sym}(\rho) I^2 + O(I^4) +..$, with $\rho=\rho_p+\rho_n$ and
 $\rho_3=\rho_p-\rho_n$ being the total and the isospin densities
and $I=\rho_3/\rho$ the isospin or asymmetry.

The predictions on $\esy(\rho)$ based on the existing many-body techniques diverge rather widely\cite{Fuchs:2005yn,Li:2008gp} both at sub-saturation and supra-saturation densities, as shown in
Fig.1 (left) for several relativistic and non-relativistic models.
Therefore, while $E_{sym}$ is fairly well constrained around saturation density from nuclear structure
its behavior at lower and at higher density is poorly known\cite{Li:2008gp,Baran:2004ih}.

The study of the $\esy$ below $\rho_0$ has been intensively pursued in the last ten years
looking at a large variety of observables at different reaction mechanisms from giant dipole resonances, pre-equilibrium emission, multi- and neck-fragmentation, isospin diffusion, etc. \cite{Li:2008gp,Baran:2004ih}.
The present contribution instead focuses on the open issues for the determination of $E_{sym}(\rho)$ in the high density region which is relevant for neutron star observables, like maximum masses, the mass-radius relation and cooling \cite{Klahn:2006ir}, for hybrid star structure and the
transition to a deconfined phase \cite{DiToro:2006pq}, and for the formation of black holes.
In view of the planned experiments at GSI (ASY-EOS), and of experiments at the future facilities  CSR/Lanzhou, FAIR/GSI, RIBF/RIKEN, FRIB/MSU\cite{Li:1999wt} several observables which are sensitive to $E_{sym}$
 to the high density region\cite{Li:2008gp,Baran:2004ih,DiToro:2006hd} have been suggested,
in particular the
$n/p$ and $^3He/t$ particle ratios and their collective flows and the isospin ratios of meson
production $\pi^-/\pi^+$ and $K^0/K^+$. We critically review these issues in this contribution.

\section{The Relativistic Structure of $\esy$ }

The determination of $\esy$ at high density $\rho> 1.5 \rho_0$ needs to exploit HIC's at beam energies
of $E/A\geq 200 MeV$ in order that the maximum density
is sufficiently high. For the theoretical approaches this shifts the energy regime of interest to a range where relativistic
effects can start to become significant. Hence relativistic, covariant approaches should be preferred;
nonetheless relativistic extensions of non-relativistic approaches represent a viable way even if they may mix effects of relativity and many-body effects\cite{Greco:2002sp}.
In this section we briefly recall the relativistic structure of $\esy$ in effective models based on nucleons
interacting through mesonic fields. At the level presented here the structure of $\esy$
is shared by the different models based on effective Lagrangians of Quantum-Hadro-Dynamics (QHD), as the Relativistic Mean Field Theory (RMFT)\cite{Greco:2000dt,Liu:2001iz}, the Density Dependent Relativistic Hartree approach (DDRH)\cite{Fuchs:1995as}, the Effective Field Theory (EFT) density functional\cite{Rusnak:1997dj} as well as the microscopic DBHF\cite{deJong:1997mn,Alonso:2003aq}.
Therefore the following discussion allows to set a common language that will be useful
in the discussion of meson production in Sect. \ref{sec:Mesons}.

In QHD the effect of nuclear interactions results
in in-medium modifications of the scalar effective masses and the energy-momentum
four-vector
\begin{equation}
\label{eff-mass}
M^{*}_{n,p}=M+\Sigma_{\sigma}\pm\Sigma_{\delta} \, \: , \: \: k^{*\mu}_{n,p}=k^{\mu}+ \Sigma^{\mu}_{\omega}\pm\Sigma^{\mu}_{\rho} \: ,
\end{equation}
where the different self-energies $\Sigma_i$ are labelled by the mesons representative of the
specific spin-isospin quantum numbers of field. The upper and lower signs refer to neutrons and protons, respectively. The isovector self-energies can be written as
\begin{equation}
\label{iso-selfene}
	\Sigma_{\delta}(n,p)= - f_{\delta} \rho_{s3} \, \: \:\:\: , \: \: \: \:\,
	\Sigma^{\mu}_{\rho}(n,p)= f_{\rho} j^{\mu}_{3} \: ,
\end{equation}
where $\rho_{s3}=\rho_{sn}-\rho_{sp}$ and $j^{\mu}_{3}=j^{\mu}_{n}-j^{\mu}_{p}$ are
the scalar isospin density and the isospin current, respectively. In the most
simple models of RMFT the coupling vertices $f_{i}$ are constant while in more sosphisticated models. as e.g. the DDHF, they are density and even momentum dependent\cite{Typel:2002ck}.

\begin{figure}[th]
\label{symm-ene}
\includegraphics[height=2.1in,width=3.4in]{Esym1.eps}
\includegraphics[height=2.2in,width=2.3in]{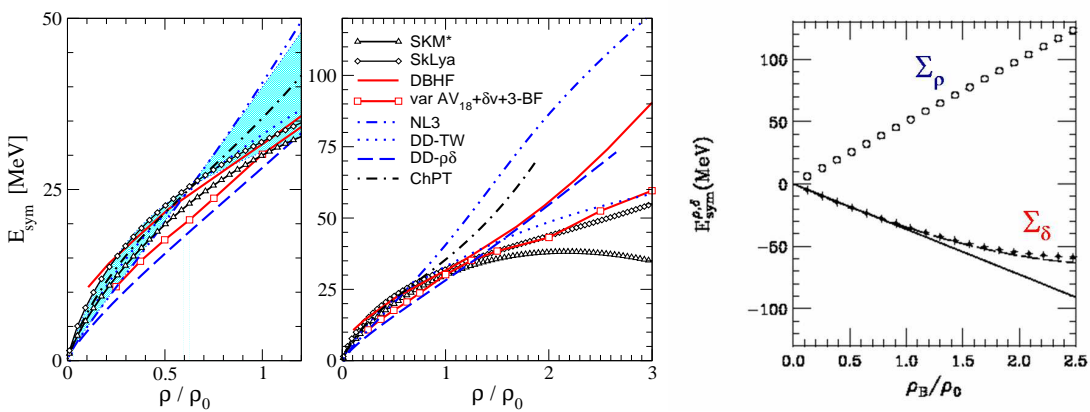}

\caption{Left and middle panel: Symmetry energy as a function of density as predicted
by different models from Ref.\cite{Fuchs:2005yn}.
Right panel: Scalar and vector isovector
self-energies in NL-RMFT; the solid line is to highlight the deviation from linearity.
}
\end{figure}

The symmetry energy arising from such a structure can be written as\cite{Liu:2001iz}
\begin{equation}
\label{esymm}
E_{sym} = \frac{1}{6} \frac{k_{F}^{2}}{E_{F}} +
\frac{1}{2} \left[ f_{\rho} - f_{\delta}\left( \frac{m^{*}}{E_F} \right)^{2} \right] \rho_{B}
\end{equation}
where $f_\rho, f_\delta$ are the couplings of the nucleons to the effective isovector scalar and vector field. However
already from the Hartree-Fock theory as well as from the more complete DBHF approach it is clear that each spin-isospin channel receives contributions from
all the meson-like fields and the meson label is only indicative of the channel\cite{Greco:2000nx}.
Eq.(\ref{esymm}) can be strictly derived only in RMFT,
but for our purpose the key point is that $\esy$ arises
from a competition of two fields, one scalar and one vector, that at $\rho \sim 2\rho_0$ are of the order
of $\sim 100$ MeV, as is shown in
Fig. 1 (right) for a standard non-linear (NL)-RMFT model
\cite{Liu:2001iz}.
This behaviour is shared by all relativistic approaches, with the difference that for example in DDRH the coupling $f_{\rho,\delta}$ depends on density, similarly in DBHF.

We stress the distinction between a scalar and a vector field because their balance in a system
in equilibrium is generally dynamically broken for an evolving system. In Ref.\cite{Greco:2002sp}
it has been shown that such an effect simulates a stiffer symmetry energy in the build-up of
collective flows. In fact, the difference of the force acting on a neutron and a proton moving with momentum
$\vec p$ can be written, after some approximations, as
\begin{equation}
\label{esym-hic}	
\frac{d\vec p_p}{d\tau}-\frac{d\vec p_n}{d\tau}\simeq 2\left[\gamma f_\rho -\frac{f_\delta}{\gamma} \right]\vec \nabla \rho_3 > \frac{4}{\rho_B} E^{pot}_{sym} \vec \nabla \rho_3 \: ,
\end{equation}
where $\gamma$ is the Lorentz factor $E/m$. We can see that dynamically the vector field is enhanced
and the scalar field suppressed resulting in an effective $\esy^*$ larger than $E^{pot}_{sym}$
as can be seen by comparing Eqs.(\ref{esymm}) and (\ref{esym-hic}).
Hence, the relativistic structure implies a modified relation between $\esy$ and observables in HIC's.
Of course, one can simulate this relativistic structure with the help of a specific
momentum dependence, but this entangles many-body effects with relativistic ones.
A first analysis has shown that such effects are significant for $E/A > 1$ GeV
\cite{Greco:2002sp}.

\section{Isospin Momentum Dependence}
\label{sec:IsospinMomentumDependence}

The many-body physics involved is indeed more complex than the one
sketched above. In fact, the RMFT class of models are derived from
a zero-range approximation that makes the meson propagator independent on the momentum.
However, the explicit momentum dependence (MD) originating in the finite range of the interaction is known
to be non-negligible in both relativistic and non-relativistic approaches. In fact, it has been shown
that the self-energies of DBHF cannot be reproduced by the simple structure of RMFT, but an explicit MD is necessary. This has lead to the development of a RMFT with derivate coupling \cite{Typel:2002ck}
that is able
to reproduce the correct energy dependence of the optical potential.
In Fig. 2 (left) the experimental optical potential is compared to the predictions of different
theoretical approaches. We can see that at high momenta $k$ also the DBHF is not able
to reproduce $U_{opt}(k)$.

The momentum dependence of the field is usually parametrized in terms of the (Landau) effective in-medium masses,
defined as
\begin{equation}
\label{effmass}
\frac{m^*_q}{m}=\left[1+\frac{m}{p}\frac{\partial U_q}{\partial p}\right] \: ,
\end{equation}
where we use the index $q=p,n$ to underline that,  in general, they are different for proton and neutrons
in asymmetric nuclear matter.
The isospin independent part of the optical model has been studied since many years and its behavior has been
fairly well determined, see
Fig. 2 (left, filled diamonds).
Instead, the isospin dependence of the optical model, usually discussed in terms of the Lane potential $U_{Lane}=(U_p-U_n)/2I$,
shown in Fig.2 (right),
 is poorly known,  and this  only in a restricted energy range, $k \leq 1 fm^{-1}$, where some data
with large uncertainties are available.

\begin{figure}[th]
\label{uopt}
\centerline{\includegraphics[height=2.3in,width=3.9in]{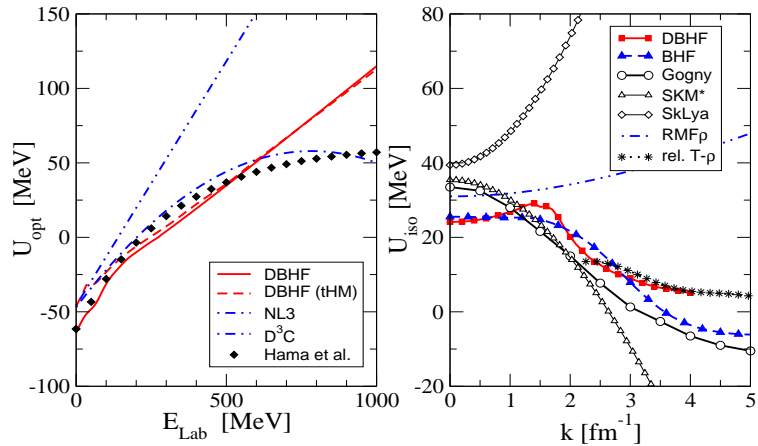}}
\vspace*{1pt}
\caption{Nucleon optical potential in nuclear matter at $\rho_0$.
On the left side DBHF calculations for symmetric nuclear matter are compared to the phenomenological
models NL3
and D3C
and to the phenomenological p-A scattering analysis (diamonds).
The right panel compares the iso-vector optical potential for several relativistic and non-relativistic
models.
Figure taken from Ref.\cite{Fuchs:2005yn}.
}
\end{figure}

We will see that all the observables under investigation are in principle affected by the MD
of the mean fields. This represents a further obstacle for the determination of $\esy$ especially
at high $\rho$,
because the importance of MD is expected to increase with energy. However, the possibility to have
access experimentally to several observables as a function of momentum in a wide range offers
the real possibility to constrain such a dependence.
Generally, all RMFT approaches give $m^{*}_{n}<m^{*}_{p}$, while in non-relativistic
approaches generally $m^{*}_{n}>m^{*}_{p}$ with some exceptions\cite{Rizzo:2003if}. Non-relativistic  BHF\cite{Bombaci:1991zz} would indicate the last choice as the correct one, which means that $U_{Lane}$ decreases with $k$.
In RMFT one finds the opposite trend but it lacks the effect of the finite range of the interaction. In DBHF both relativistic fields and the finite range effect are included, but there exists an ambiguity from the method used to project on the different Lorentz amplitudes\cite{Fuchs:2005yn}.
One could judge that the microscopic approaches favour the relation $m^{*}_{n}>m^{*}_{p}$, but
in any case it is mandatory to show that this is the case in HIC
by means of a comparison with the available and forthcoming data.

\begin{figure}[th]
\label{np-ratio}
\includegraphics[height=2.3in,width=5.8in]{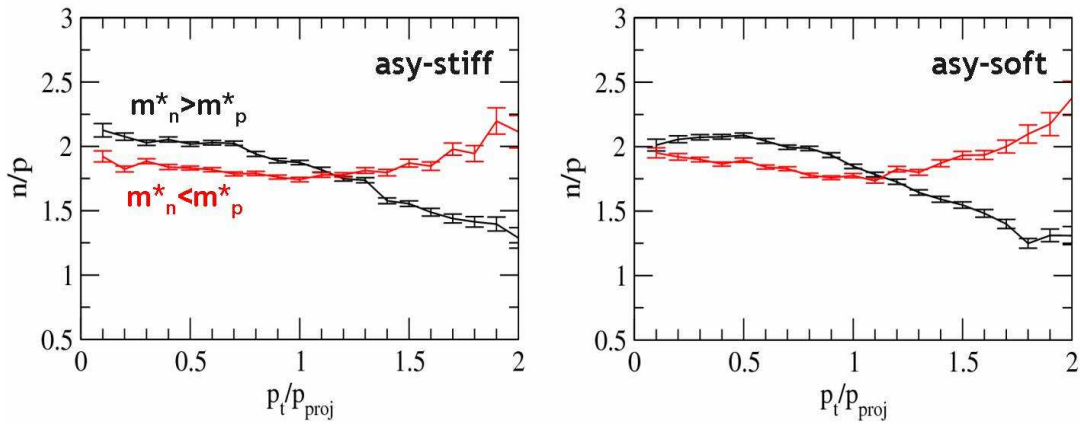}

\vspace*{1pt}
\caption{n/p ratio for $^{132}Sn+^{124}Sn$ at 400 AMeV (b=1fm, $y < 0.3$) vs. the the transverse momentum normalized
to the projectile one for two choices of $\esy(\rho)$ and $U_{Lane}$.}
\end{figure}

Investigations of the MD of the isovector interaction have been performed
using a non-relativistic transport code (Boltzmann-Nordheim-Vlasov, BNV) with a Skyrme-like mean field modified to account for a different MD for protons and neutrons\cite{Rizzo:2003if}.
At $E/A= 400$ AMeV for central collisions one can reach densities of about $1.8 \rho_0$
for a time interval of the order of 15-20 fm/c.
More interestingly, it is possible to find a kinematical region ($y\leq 0.3$) where the
$n/p$ and $^3H/^3He$ ratios of emitted particles as a function of the transverse momentum depend mainly on the neutron-proton splitting of the effective masses and are nearly independent of
the stiffness of $\esy$.
In Fig. 3 one sees that the $n/p$ ratio exhibits a strong dependence
for $p_t\geq p_{proj}=2 fm^{-1}$ depending on the sign of the splitting of the effective masses. For $m^{*}_{n}<m^{*}_{p}$ the difference between the neutron and proton mean fields increases with momentum
and the more energetic neutrons feel an increasing repulsion resulting in a larger
$n/p$ ratio.
The opposite is true for the case $m^{*}_{n}<m^{*}_{p}$. The key point of our result \cite{DiToro:2008zm} is that especially
at high $p_t$ the effect is almost independent on the stiffness of the symmetry energy allowing
to disentangle the dependence on the momentum from that on the density.
Instead, for lower beam energies and in the low momentum region the pre-equilibrium emission is found to
depend significantly on both the stiffness of $\esy$ and the isospin MD.

\section{Isopin effects on Meson Production}
\label{sec:Mesons}

The study of the symmetry energy at high density implies to investigate HIC's at higher energy
where a system at sufficiently high density is created. On the other hand, then the energy becomes large enough to excite the lower mass mesonic states,
namely pions and kaons. Therefore symmetry energy effects are transferred to the production of mesons.
In Ref.\cite{Li:2002xq} it was suggested that the $\pi^-/\pi^+$ ratio is sensitive to $\esy(\rho)$. It was demonstrated that the main effect consists in the
emission of nucleons during the evolution depending on $\esy$ leading to a different $n/p$ content of the residual system, which in turn influences the formation of the different charge states of the pion.

\begin{figure}[th]
\label{ratio-time}
\centerline{\includegraphics[height=2.2in,width=3.8in]{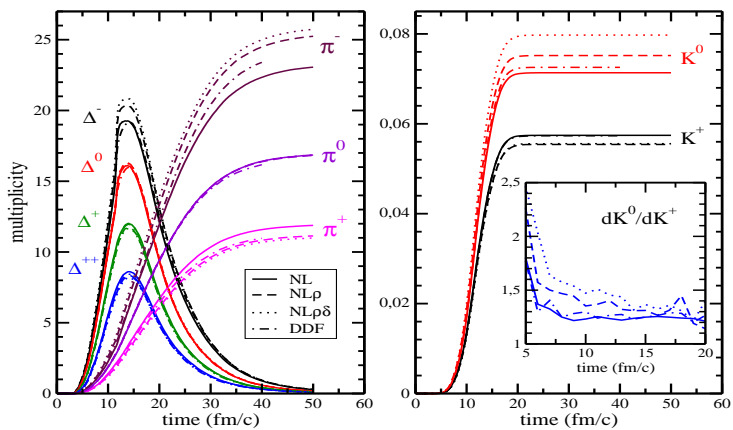}}
\vspace*{1pt}
\caption{Time evolution of the $\Delta^{\pm,0,++}$ resonances and pions $\pi^{\pm,0}$ (left),
and  of kaons ($K^{+,0}$(right) for a central ($b=0$ fm impact parameter)
Au+Au collision at 1 AGeV incident energy. Transport calculation using the
$NL, NL\rho, NL\rho\delta$ and $DDF$ models for the iso-vector part of the
nuclear $EoS$ are shown. The inset shows the differential $K^0/K^+$  ratio
as a function of the kaon emission time.}
\end{figure}

Even if this mechanism seems quite straightforward one may doubt its real effectiveness. Indeed, the idea of
using pions to determine the symmetric part of the EOS has suffered from the fact that pions
strongly interact with nucleons and are produced during the whole evolution of the collision system making it difficult to associate
their production to a specific density reached during the collision.
In that context it was suggested by Aichelin and Ko\cite{Aichelin:1986ss} that kaons are a better probe of the EOS. The reason is manifold: Kaons have a higher threshold energy, hence they are produced only in the
high density phase. Moreover, once produced they interact weakly with nucleons and their width with respect
to the mass is quite small making a quasi-particle approximation more reliable.
After nearly 20 years the effort to determine the symmetric EOS by kaon production has been successfull and is summarised
in Ref.\cite{Fuchs:2005zg}.
Following the same line of thinking the Catania group has suggested to investigate the $K^0/K^+$ ratio
 as a better probe of the $\esy$ at high density \cite{Ferrini:2005jw,Ferini:2006je}
(the other isospin pair with anti-kaons $\bar{K}_0/\bar{K}^-$ suffers from the strong
coupling to the medium).

Using a relativistic transport approach (Relativistic Boltzmann-\"Uhling-Uhlenbeck, RBUU\cite{Fuchs95}) we have analyzed
pion and kaon production in central $^{197}Au+^{197}Au$ collisions in
the $0.8-1.8~AGeV$ beam energy range, comparing models with the same ``soft'' EOS for symmetric
matter and with different effective field choices for $E_{sym}$.
Fig. 4 reports  the temporal evolution of $\Delta^{\pm,0,++}$
resonances, pions ($\pi^{\pm,0}$) and kaons ($K^{+,0}$)
for central Au+Au collisions at $1AGeV$\cite{Ferini:2006je,Prassa07}
It is clear that, while the pion yield freezes out at times of the order of
$50 fm/c$, i.e. at the final stage of the reaction (and at low densities),
kaon production occurs within the very early (compression) stage,
 and the yield saturates at around $15 fm/c$, when the nucleon
and $\Delta's$ densities reach their maximum value.
From Fig. 4 we see that the pion multiplicities are
moderately dependent on the isospin part of the nuclear mean field.
However, a slight increase (decrease) in the $\pi^{-}$ ($\pi^{+}$)
multiplicity is observed when going from a
NL-RMFT model without isovector fields (NL) to one  with a $\rho$ meson ($NL\rho$) and with $\rho$ and $\delta$ mesons ($NL\rho\delta$).
 This trend is more pronounced for kaons, see the
right panel, due to the high density selection of the source and the
proximity to the production threshold.
In Fig. 5 (left) the pion and kaon rations from these calculations are shown as a function of beam energy. The ratios decrease with beam energy, because the relative effects of mean fields and thresholds become less important. However, the greater sensitivitiy of the kaon ratios is seen clearly.

\begin{figure}[th]
\label{excitation}
\includegraphics[height=2.3in,width=2.6in]{kratio-ene.eps}
\hspace*{10pt}
\includegraphics[height=2.3in,width=2.8in]{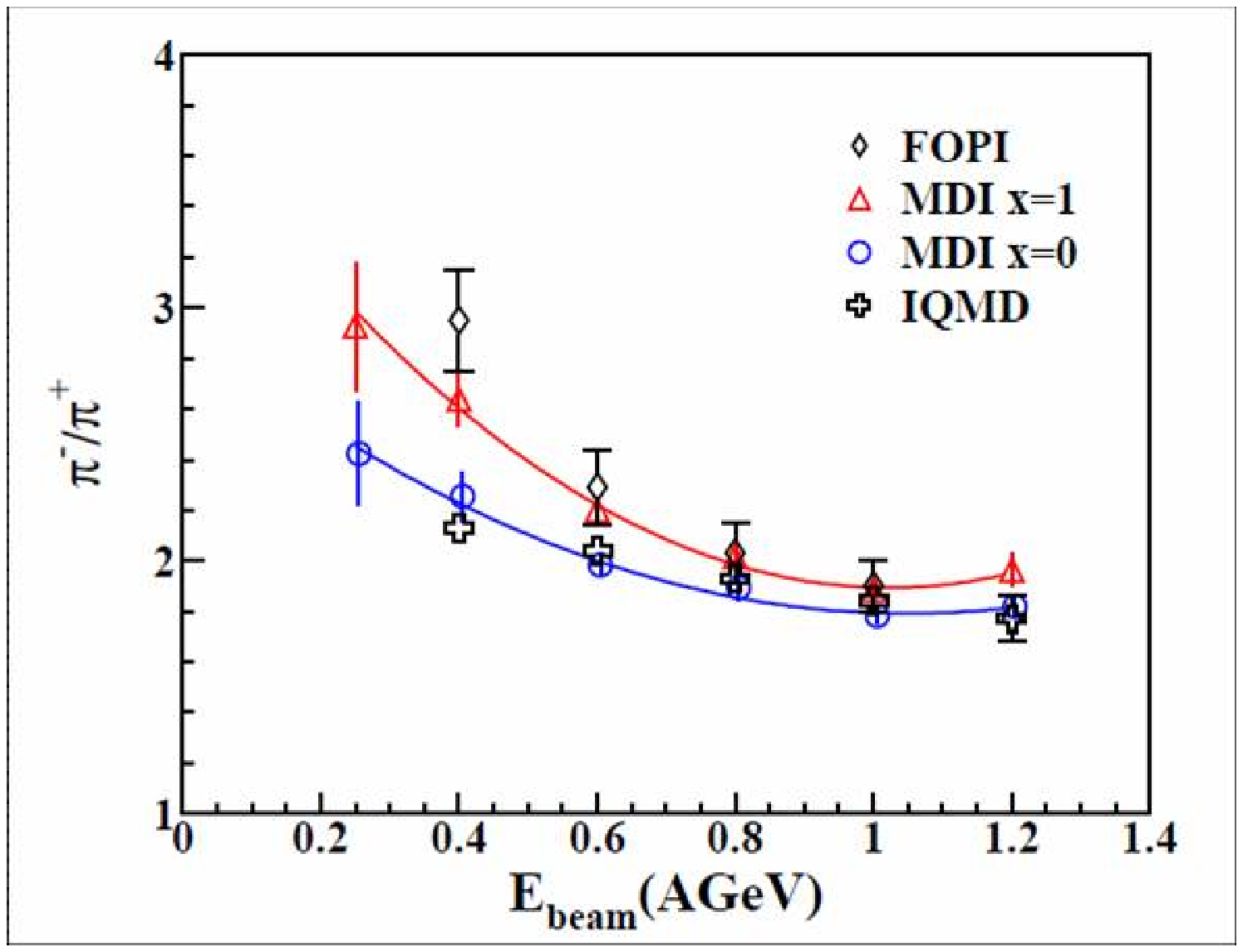}
\vspace*{1pt}
\caption{Left: Excitation function of the $\pi^-/\pi^+$ and $K^0/K^+$ ratios for $Au+Au$. RBUU results for different
behavior of $\esy(\rho)$\cite{Ferini:2006je}. Right: Results for $Ru+Ru$ in IBUU04 for a soft (x=1) and a stiff (x=0) $\esy$\cite{Xiao:2009zza}
compared to the data from FOPI and a calculation with IQMD\cite{Reisdorf:2006ie}.
}
\end{figure}

At present there are essentially no data for kaons, while there are some for pions and $\pi^-/\pi^+$ ratios from
FOPI collaboration \cite{Reisdorf:2006ie}. Within an isospin and momentum dependent transport model
it has been shown that an agreement with data can be achieved only if a very soft $\esy(\rho)$ is
employed \cite{Xiao:2009zza}. Such a finding appears in strong disagreement with other studies
that exploit the elliptic flow to extract the slope of $\esy$ around and above the saturation density \cite{Trautmann:2009kq}.
However there are circumstantial reasons to be careful. Indeed, the physics
involved in the in-medium particle production has many aspects and a comprehensive and self-consistent
approach is necessary before extracting the isospin dependence of the interaction.

The effect described in Ref.\cite{Xiao:2009zza} essentially is due to the fact that a stiff
$\esy$ causes a neutron-rich emission of nucleons in the early stages of the reaction leaving the system
too symmetric in isospin content to reproduce the $\pi^-/\pi^+$ ratio of FOPI.
On the other hand, if one employs an $\esy(\rho)$ that just above $\rho_0$ decreases with density generating an isospin
force that is attractive for the neutrons it is possible to get close to the data,
see Fig. 5 (right).
However, we note that it is mandatory to check if one can reproduce at the same time the
$n/p$ emission as a function of $p_t$ especially at the high $p_t$ relevant for pion production, since this represents the complementary signal for the soft  $\esy$.
This would be a first test of the claim for a very soft $E_{sym}$ and
in case of failure it will
provide evidence that there are other mechanism determining the in-medium
meson production.
Indeed, we already know that there are at least other three effects competing
with the mean field effect on the $n/p$ emission.
These are the so-called "`threshold effect"'
emphasized by the calculation of Ref.s \cite{Ferrini:2005jw,Ferini:2006je},
the momentum dependence of the isospin-dependent cross sections, and the
 the isospin mass shift of pions due to the coupling to $\Delta-$hole excitation\cite{Xu:2009fj}. These will be discussed in the following.

\subsection{Isospin dependence of thresholds and spectral functions}
\label{sec:threshold}

The "`threshold effect"' is due to the fact that the masses of nucleons and
Delta's are modified in the medium.
The unknown self-energies of the delta's are usually specified in terms of the neutron and proton ones by the use
of Clebsch-Gordon coefficients for the isospin coupling of the $\Delta's$ to nucleons
\cite{Li:2002xq,Ferrini:2005jw}.
These medium modifications are isospin dependent, Eq.(\ref{iso-selfene}).
This influences the phase-space available for meson production in a nucleon-nucleon (NN) collision because it modifies the difference between the invariant energy
in the entrance channel $s_{in}$ with respect to the production threshold $s_{th}$.
This effect is, of course,
present in general for all meson productions, but for brevity we concentrate here on
one inelastic channel: $nn\rightarrow p\Delta^-$, mainly responsible for $\pi^-$ production.
From Eq.(\ref{eff-mass}) the invariant energy in the entrance channel and the threshold energy are given, respectively, by

\begin{eqnarray}
\label{threshold}
\sqrt{s_{in}}/2= \left[ E^{*}_{n}+\Sigma^{0}_{n}\right]\stackrel{p=0}{\rightarrow}\left[m^{*}_{N}
+\Sigma^{0}_{\omega}+\Sigma^{0}_{\rho}+\Sigma_{\delta}\right]>m^{*}_{N}+\Sigma^{0}_{\omega} \nonumber\\
\sqrt{s_{th}}= \left[m^{*}_{p}+m^{*}_{\Delta^-}+\Sigma_{0}(p)+\Sigma_{0}(\Delta^-)\right]=
\left[m^{*}_{N}+m^{*}_{\Delta}+2 \Sigma_{\omega} \right]
\end{eqnarray}

where $m^{*}_{N}, m^{*}_{\Delta}$ are the isospin averaged values.
The last equality for the threshold energy is valid due to the prescription for the Delta self-energies noted above, which leads to an exact compensation of the isospin-dependent parts, hence
the threshold $s_{th}$ is not modified by isospin dependent self-energies.
In generally in a self-consistent many-body calculation higher order effects can destroy this exact balance.
But this is not so important for our qualitative discussion, since there will always be a compensating effects in $s_{th}$.
On the other hand, the energy available in the entrance channel, $s_{in}$, is
shifted in an explicitely isospin dependent way by the in-medium self-energy $\Sigma^{0}_{\rho}+\Sigma_{\delta}>0$.
Especially, the vector self-energy gives a positive
contribution to neutrons that increases the difference $s_{in}-s_{th}$ and hence increases the cross section of the
inelastic process due to the opening up of the phase-space, expecially close to threshold.


A similar modification but opposite in sign is present in $s_{in}-s_{th}$  for the $pp\rightarrow n\Delta^{++}$
channel that therefore is suppressed by the isospin effect on the self-energies.
Hence, due to the described threshold effect the ratio $\pi^{-}/\pi^{+}$ increases with the stiffness
of $\esy$ which is associated with a large $\Sigma^0_{\rho}+\Sigma_{\delta}$. This is at the
origin of the result in
Fig. 5 (left). Of course, the RBUU calculation contains also
the mean field effect on the isospin content but the final result appears to be dominated by the
threshold effect. In fact, it is seen that $\pi^{-}/\pi^{+}$ increases with the increasing stiffness of $\esy$ and the effect is stronger closer to the threshold energy.

The threshold effect described constitutes one physical effect that is present in the in-medium
meson production and results in an opposite dependence on $\esy$ relative to the mean field effect on
the emission of neutrons and protons. On the other hand this does not mean that in Ref.\cite{Ferini:2006je}
the effect is quantitatively described in the more realistic way.
There are at least two other physical aspects that have to be considered.
One is the fact that the self-energies in the implementation of RBUU is not explicitly momentum
dependent as it is in DBHF and as the study of the optical potential shows,
see Fig. 2 (right).
Thus we return to the importance of the isospin momentum dependence discussed in
Sect. \ref{sec:IsospinMomentumDependence}.
Because we neglect the dependence of the self-energies on momentum,
the difference $s_{in}-s_{th}$ in Eq.(\ref{threshold}) increases strongly with the momentum
$p$. The problem is directly related to the strong energy dependence of the optical
potential in RMFT in Fig. 2.
Therefore it is likely that a more realistic calculation
which includes a $\Sigma_{\rho,\delta}({\rho,p})$ will reduce the effect seen in
Fig. 5.
In addition a fully consistent treatment should include an optical potential
also for the pions.
It can be envisaged that a stronger sensitivity to $\Sigma({\rho,p})$ can be seen
looking at $\pi^{-}/\pi^{+}$ as a function of the pion transverse energy.

Another key point in the study of the isospin effects is the isospin
dependence of the cross
section for the inelastic processes. In fact the same medium effects that produce
the threshold effect discussed above also modify the cross section of the pertinent inelastic processes.
The problem of the self-consistency between the mean field and the cross section
is known for many years, however, in the case under discussion it seems to be much more
relevant. In fact
one can simulate the effect of a repulsive mean field, e.g. on collective
flow, by increasing the cross section.
However, if one plays with the cross section for inelastic collisions it will manifest itself in the number of particles that are produced and measured.
In this respect the calculations of pion and kaon yields of Ref.\cite{Ferini:2006je} take
into account only one consequence of medium effects, namely the isopin dependent self-energies. It is likely
that simultaneously taking into account the isospin dependence of the cross section
can dampen the effect of self-energies on $s_{in}-s_{th}$.

Another effect that is certainly present is the modification
of the pion spectral function which in an asymmetric medium becomes isospin dependent.
The effect has been pointed out in Ref. \cite{Xu:2009fj}, see also the contribution of C.M. Ko in these Proceedings. This work shows that the
increased strength at low energies  of the $\pi^-$ relative to the $\pi^+$ spectral function in dense asymmetric nuclear matter
modifies the expected thermal $\pi^-/\pi^+$ ratio with respect to the one with vacuum masses.
The effect is comparable in size to the mean field effect on $n/p$ ratios of emitted nucleons,
but again goes in the opposite direction: larger $E_{sym}$ implies larger $\pi^-/\pi^+$ ratios.
Therefore it would be important to include such an effect in the transport models,
even if it means to go beyond the simple quasi-particle approximation.
From this point of view it would be desirable
to have calculations with transport codes including off-shell effects\cite{Cassing:2008nn}.

Considering these various issues that are involved in the determination of the
$\pi^-/\pi^+$ as well as the $K^0/K^-$ ratios the circumstantial evidence
for a very soft $\esy$ at high density should be considered only as a first step
that opened the discussion. Instead, there is a need for theoretical
improvements in the transport models to match the upcoming availability of
experiments able to scrutiny more precisely the isopin dependence of observables in heavy ion collisions.

\section{Outlook}
\label{sec:Outlook}
The determination of $\esy(\rho)$ above the saturation density by means of
HIC's is indeed at the beginning.
With increasing energy the momentum dependence of the mean field determines the
subsequent dynamics. Therefore the study of the $\esy$ is entangled with
the isospin effects on the momentum dependence. This affects
not only the nucleon dynamics but also the isopin content of the meson production.
We discussed several issues which from a theorethical point of view require further development of the present transport theories
for the study of the isospin dependence of meson production.
From the experimental point of view the possibility to look at observables
especially as a function of momentum should allow to really constrain the $\esy$
in the high density region. However a key point of the search for the high-density symmetry energy  should be to look at several observables in the same reaction simultaneously which should help not to
put under the carpet the important mechanisms at work in HIC dynamics.

\section*{Acknowledgements}
V. Greco would like to thank the organizers of IWDN09 for their warm hospitality
and T. Gaitanos for useful comments on the preparation of the talk.


\section*{References}
\begin{thebibliography}{10}

\bibitem{Danielewicz:2002pu}
P. Danielewicz, R. Lacey and W.G. Lynch,
\newblock Science 298 (2002) 1592, nucl-th/0208016.

\bibitem{Fuchs:2005yn}
C. Fuchs and H.H. Wolter,
\newblock Eur. Phys. J. A30 (2006) 5, nucl-th/0511070.


\bibitem{Li:2008gp}
B.A. Li, L.W. Chen and C.M. Ko,
\newblock Phys. Rept. 464 (2008) 113, 0804.3580.

\bibitem{Baran:2004ih}
V. Baran et~al.,
\newblock Phys. Rept. 410 (2005) 335, nucl-th/0412060.

\bibitem{deJong:1997mn}
F. de~Jong and H. Lenske,
\newblock Phys. Rev. C57 (1998) 3099, nucl-th/9707017.

\bibitem{Klahn:2006ir}
T. Klahn et~al.,
\newblock Phys. Rev. C74 (2006) 035802, nucl-th/0602038.

\bibitem{DiToro:2006pq}
M. Di~Toro et~al.,
\newblock Nucl. Phys. A775 (2006) 102, nucl-th/0602052.

\bibitem{Li:1999wt}
X.D. Li et~al.,
\newblock Phys. Rev. Lett. 83 (1999) 3776, hep-ph/9905356.

\bibitem{DiToro:2006hd}
M. Di~Toro et~al.,
\newblock Nucl. Phys. A787 (2007) 585, nucl-th/0609081.

\bibitem{Greco:2002sp}
V. Greco et~al.,
\newblock Phys. Lett. B562 (2003) 215, nucl-th/0212102.

\bibitem{Greco:2000dt}
V. Greco et~al.,
\newblock Phys. Rev. C64 (2001) 045203, nucl-th/0011033.

\bibitem{Liu:2001iz}
B. Liu et~al.,
\newblock Phys. Rev. C65 (2002) 045201, nucl-th/0112034.

\bibitem{Fuchs:1995as}
C. Fuchs, H. Lenske and H.H. Wolter,
\newblock Phys. Rev. C52 (1995) 3043, nucl-th/9507044.

\bibitem{Rusnak:1997dj}
J.J. Rusnak and R.J. Furnstahl,
\newblock Nucl. Phys. A627 (1997) 495, nucl-th/9708040.

\bibitem{Alonso:2003aq}
D. Alonso and F. Sammarruca,
\newblock Phys. Rev. C67 (2003) 054301, nucl-th/0301032.

\bibitem{Typel:2002ck}
S. Typel, T. van Chossy and H.H. Wolter,
\newblock Phys. Rev. C67 (2003) 034002, nucl-th/0210090;
S. Typel, Phys. Rev. C71 (2005) 064301, nucl-th/0501056

\bibitem{Greco:2000nx}
V. Greco et~al.,
\newblock Phys. Rev. C63 (2001) 035202, nucl-th/0010092.

\bibitem{Rizzo:2003if}
J. Rizzo et~al.,
\newblock Nucl. Phys. A732 (2004) 202, nucl-th/0309032.

\bibitem{DiToro:2008zm}
M. Di~Toro et~al.,
\newblock Prog. Part. Nucl. Phys. 62 (2008) 389, 0811.2880.

\bibitem{Bombaci:1991zz}
I. Bombaci and U. Lombardo,
\newblock Phys. Rev. C44 (1991) 1892.

\bibitem{Li:2002xq}
B.A. Li,
\newblock Phys. Rev. C67 (2003) 017601, nucl-th/0211037.

\bibitem{Aichelin:1986ss}
J. Aichelin and C.M. Ko,
\newblock Phys. Rev. Lett. 55 (1985) 2661.

\bibitem{Fuchs:2005zg}
C. Fuchs,
\newblock Prog. Part. Nucl. Phys. 56 (2006) 1, nucl-th/0507017.

\bibitem{Ferrini:2005jw}
G. Ferini et~al.,
\newblock Nucl. Phys. A762 (2005) 147, nucl-th/0504032.

\bibitem{Ferini:2006je}
G. Ferini et~al.,
\newblock Phys. Rev. Lett. 97 (2006) 202301, nucl-th/0607005.

\bibitem{Fuchs95}
C. Fuchs, H.H. Wolter, Nucl. Phys. A589 (1995) 732.

\bibitem{Prassa07}
V. Prassa, et~al., Nucl. Phys. A789 (2007) 311.

\bibitem{Xiao:2009zza}
Z. Xiao et~al.,
\newblock Phys. Rev. Lett. 102 (2009) 062502.

\bibitem{Reisdorf:2006ie}
W. Reisdorf et~al. (FOPI collaboration),
\newblock Nucl. Phys. A781 (2007) 459.

\bibitem{Trautmann:2009kq}
W. Trautmann et~al.,
\newblock Prog. Part. Nucl. Phys. 62 (2009) 425, 0904.3495.

\bibitem{Xu:2009fj}
J. Xu, C.M. Ko and Y. Oh,
\newblock (2009), 0906.1602.

\bibitem{Cassing:2008nn}
W. Cassing,
\newblock Eur. Phys. J. ST 168 (2009) 3, 0808.0715.

\end{thebibliography}
\end{document}